\documentclass[10pt]{iopart}
\usepackage{epsfig,cite}
\def\lsim{\raise0.3ex\hbox{$<$\kern-0.75em\raise-1.1ex\hbox{$\sim$}}}
\def\gsim{\raise0.3ex\hbox{$>$\kern-0.75em\raise-1.1ex\hbox{$\sim$}}}

\newcommand{\bec}{\begin{center}}
\newcommand{\enc}{\end{center}}
\newcommand{\beq}{\begin{equation}}
\newcommand{\eeq}{\end{equation}}
\newcommand{\ds}{\displaystyle}
\newcommand{\beqar}{\begin{eqnarray}}
\newcommand{\eeqar}{\end{eqnarray}}

\begin{document}

\title{
EOS at FAIR energies and the role of resonances }

\author{
E~E~Zabrodin\dag\ddag, I~C~Arsene\dag, J~Bleibel\S, 
M~Bleicher$\Vert$, L~V~Bravina\dag, G~Burau$\Vert$, Amand~Faessler\S, 
C~Fuchs\S, M~S~Nilsson\dag, K~Tywoniuk\P, H~St\"ocker$\Vert$ $^+$ *  
}  
\address{\dag\
         Department of Physics, University of Oslo, Oslo, Norway}
\address{\ddag\
         Institute for Nuclear Physics, Moscow State University,
         Moscow, Russia}
\address{\S\
         Institute for Theoretical Physics, University of T\"ubingen,
         T\"ubingen, Germany}
\address{$\Vert$\
         Institute for Theoretical Physics, University of Frankfurt,
         Frankfurt am Main, Germany}
\address{\P\
         Departamento de F{\'\i}sica de Part{\'\i}culas, Universidad de
         Santiago de Compostela, Santiago de Compostela, Spain} 
\address{$^+$\
         Gesellschaft f\"ur Schwerionenforschung mbH, Darmstadt, 
         Germany}
\address{*
         Frankfurt Institute for Advanced Studies (FIAS), University 
         of Frankfurt, Frankfurt am Main, Germany}

\begin{abstract}
Two microscopic models, UrQMD and QGSM, are used to extract the 
effective equation of state (EOS) of locally equilibrated nuclear 
matter produced in heavy-ion collisions at energies from 11.6 AGeV 
to 160 AGeV.  Analysis is performed for the fixed central cubic cell 
of volume $V = 125$\,fm$^3$ and for the expanding cell that followed 
the growth of the central area with uniformly distributed energy.
For all reactions the state of local equilibrium is nearly approached
in both models after a certain relaxation period. The EOS has a simple 
linear dependence $P = c_s^2\, \varepsilon$ with 
$0.12 \leq c_s^2 \leq 0.145$. Heavy resonances are shown to be 
responsible for deviations of the $c_s^2(T)$ and $c_s^2(\mu_{\rm B})$ 
from linear behavior. In the $T$-$\mu_{\rm B}$ and $T$-$\mu_{\rm S}$ 
planes the EOS has also almost linear dependence and demonstrates 
kinks related not to the deconfinement phase transition but to 
inelastic freeze-out in the system.
\end{abstract}




\section{Introduction}
\label{sec1}

One of the principle questions of the Compressed Baryon Matter (CBM)
experiment at GSI FAIR is the equation of state of hot and dense 
matter produced in heavy-ion collisions at energies about 20 - 40 
AGeV \cite{fair}. Because the perturbative quantum chromodynamics 
(pQCD) is not applicable to soft processes with small momentum 
transfer, one has to rely on microscopic models that correctly 
describe many features of the collisions at various energies. Two
of such models, ultra-relativistic quantum molecular dynamics (UrQMD) 
\cite{urqmd} and quark-gluon string model (QGSM) \cite{qgsm}, are used 
to extract the effective EOS of the excited matter in heavy-ion
collisions at bombarding energies ranging from AGS to SPS. The UrQMD 
and, to a lesser extent, QGSM were already employed for studying the 
equilibration processes, see \cite{urqmd_equil,qgsm_equil}. 
Recently we modified the analysis by extending it to a non-fixed cell, 
which should follow the expanding area of uniformly distributed energy 
density \cite{prc_08}. By using both the UrQMD and QGSM for 
studies of the relaxation process in a broad energy range one can 
expect that the model-dependent effects, caused by application of a 
particular event generator, will be significantly reduced. - The 
models use different mechanisms of string excitation and fragmentation. 
UrQMD relies on the longitudinal excitation, whereas the color exchange 
scheme is employed in QGSM.  The fragmentation functions that determine 
the energy, momentum, and the type of the hadrons produced during the 
string decay are also different. Finally, both models do 
not use the same tables of hadrons, chosen as discrete degrees of 
freedom. Whereas the UrQMD contains 55 baryon and 32 meson states 
together with their antistates, the QGSM takes into account octet and 
decuplet baryons, and nonets of vector and pseudoscalar mesons, as well 
as their antiparticles. Heavy resonances are not included in the 
current version of the QGSM, and this circumstance can be used to 
elaborate on their role in the EOS.

Central gold-gold collisions with zero impact parameter $b = 0$\,fm 
were simulated at bombarding energies $E_{\rm lab} = 11.6, 20, 30, 40, 
80$ and 160\, AGeV, respectively. The total energy, the net baryon 
charge and the net strangeness extracted for a certain volume of the 
reaction, were inserted into a system of nonlinear equations 
\cite{urqmd_equil} to obtain temperature $T$, baryon chemical 
potential $\mu_{\rm B}$ and strangeness chemical potential 
$\mu_{\rm S}$ of an ideal hadron gas in equilibrium. If the yields and 
transverse momentum spectra of particles obtained in a snapshot of
microscopic simulations at time $t$ were close to the results of 
statistical model (SM), the matter in the cell is considered to be 
in the vicinity of equilibrium. Then its equation of state can be 
derived and studied. Because the cell is an open system with instantly 
changing energy and particle density, the verification of the 
equilibrium conditions is repeated after a time-step of 
$\Delta t = 1$\,fm/$c$.  

\section{Relaxation to equilibrium and EOS in the cell.}
\label{sec2}

\begin{figure}[htb]
\begin{minipage}[t]{70mm}
\epsfig{file=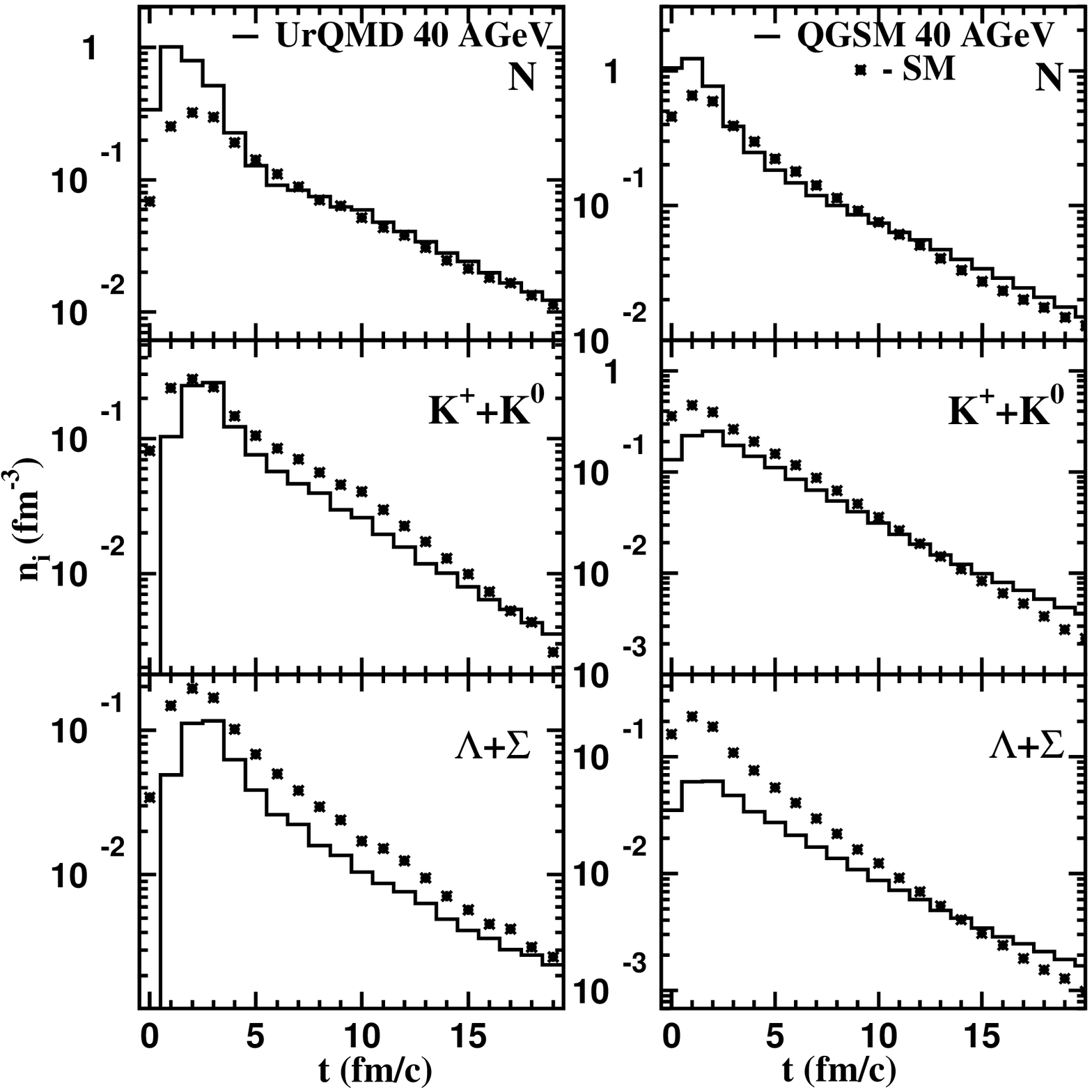,width=70mm} 
\caption{
Hadron yields in the central $V = 125$\,fm$^3$ cell
of central Au+Au collisions at 40\,AGeV in microscopic models
(histograms) and statistical model (symbols). 
}
\label{fig1}
\end{minipage}
\hspace{\fill}
\begin{minipage}[t]{65mm}
\epsfig{file=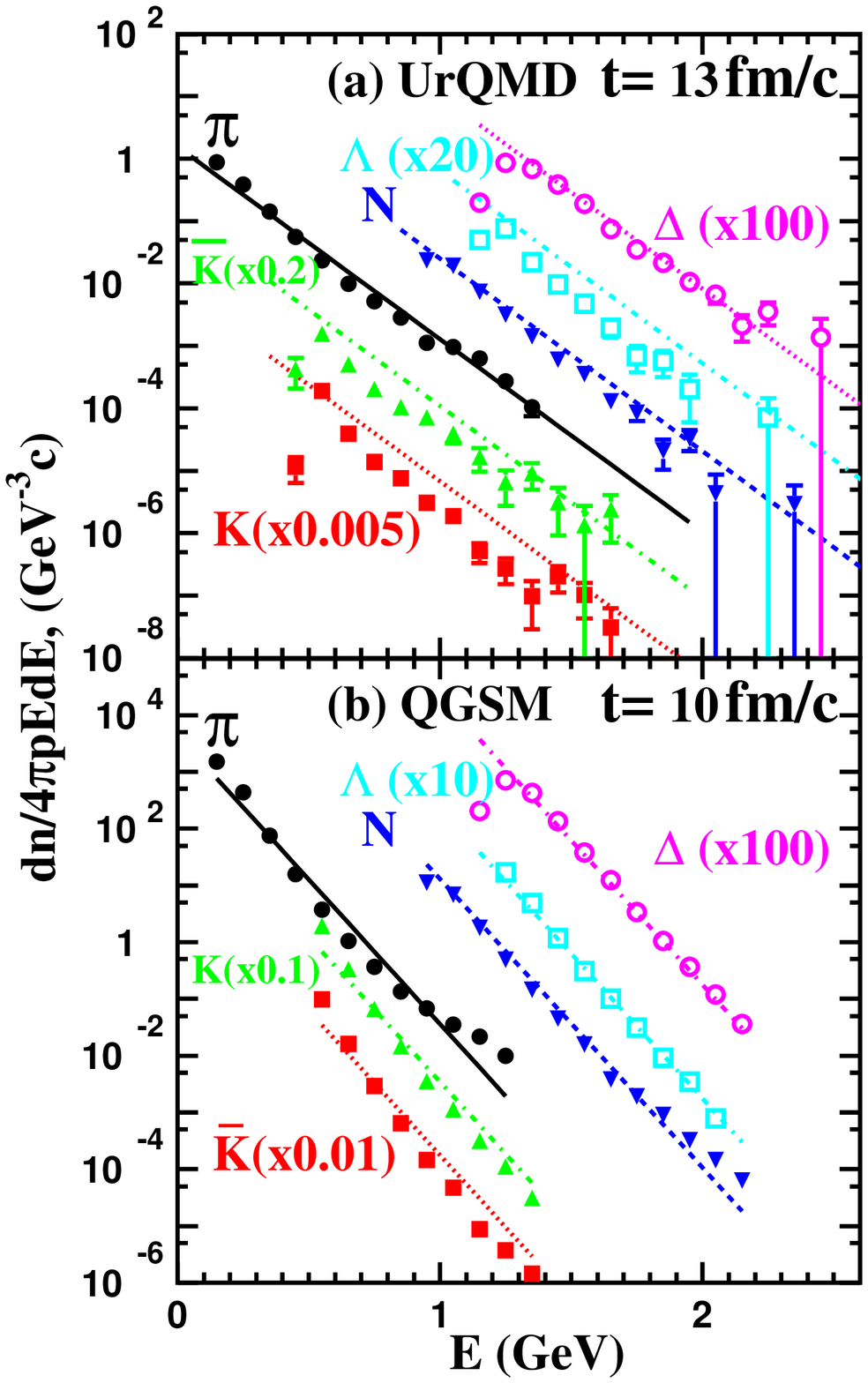,width=70mm}
\caption{
Energy spectra of hadrons in the central $V = 125$\,fm$^3$ cell. 
}
\label{fig2}
\end{minipage}
\end{figure}

In the standard approach the test-volume was a fixed central cubic
cell of $V = 125$\,fm$^3$.
The yields of some hadron species are displayed in Fig.~\ref{fig1}
for central gold-gold collisions at $E_{\rm lab} = 40$\,AGeV. The
agreement between the results of microscopic and statistical model
calculations is good after $t \geq 9$\,fm/$c$. Here the standard 
criterion $[yield(mic)-yield(SM)]/error(SM) \leq 1$ is applied.
According to model analysis, after $t \approx 10$\,fm/$c$ almost
all many-body processes going via the formation of strings or 
many-particle decaying resonances are ceased, and one deals 
mainly with elastic and quasi-elastic reactions. The energy spectra 
$d N / 4 \pi p E d E$ calculated microscopically are shown in
Fig.~\ref{fig2}. The Boltzmann fit to particle distributions is 
presented in Fig.~\ref{fig2} as well. Both in UrQMD and in QGSM the 
energy spectra agree well with the exponential form of the Boltzmann
distributions. Because the hadronic matter in the central cell nearly
reaches the state of thermal and chemical equilibrium, the macroscopic
thermodynamic parameters of the system, such as temperature and
chemical potentials, become meaningful.

\begin{figure}[htb]
\begin{minipage}[t]{70mm}
\epsfig{file=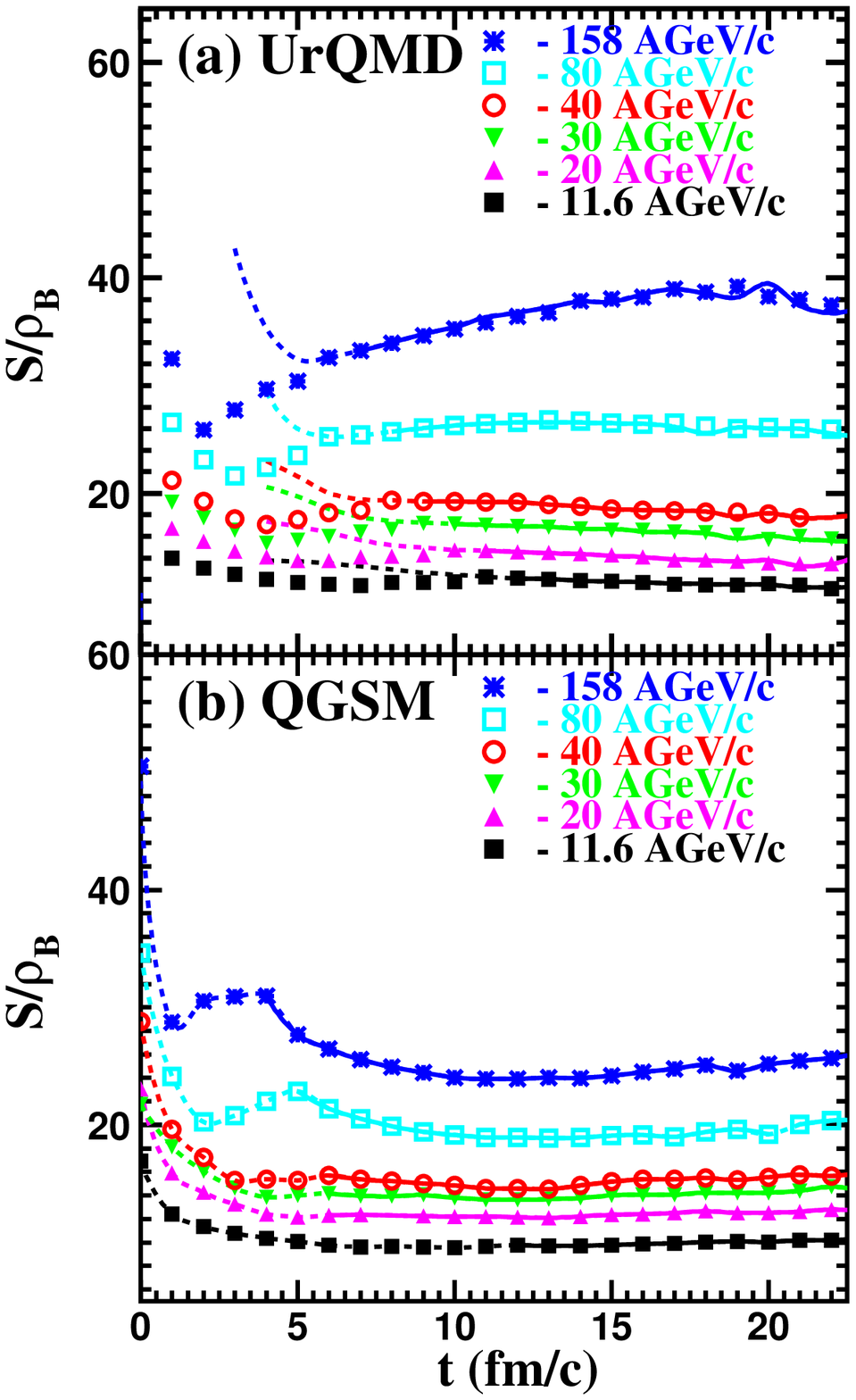,width=75mm}
\caption{
Entropy per baryon in the central cell as a function of time $t$.
}
\label{fig3}
\end{minipage}
\hspace{\fill}
\begin{minipage}[t]{70mm}
\epsfig{file=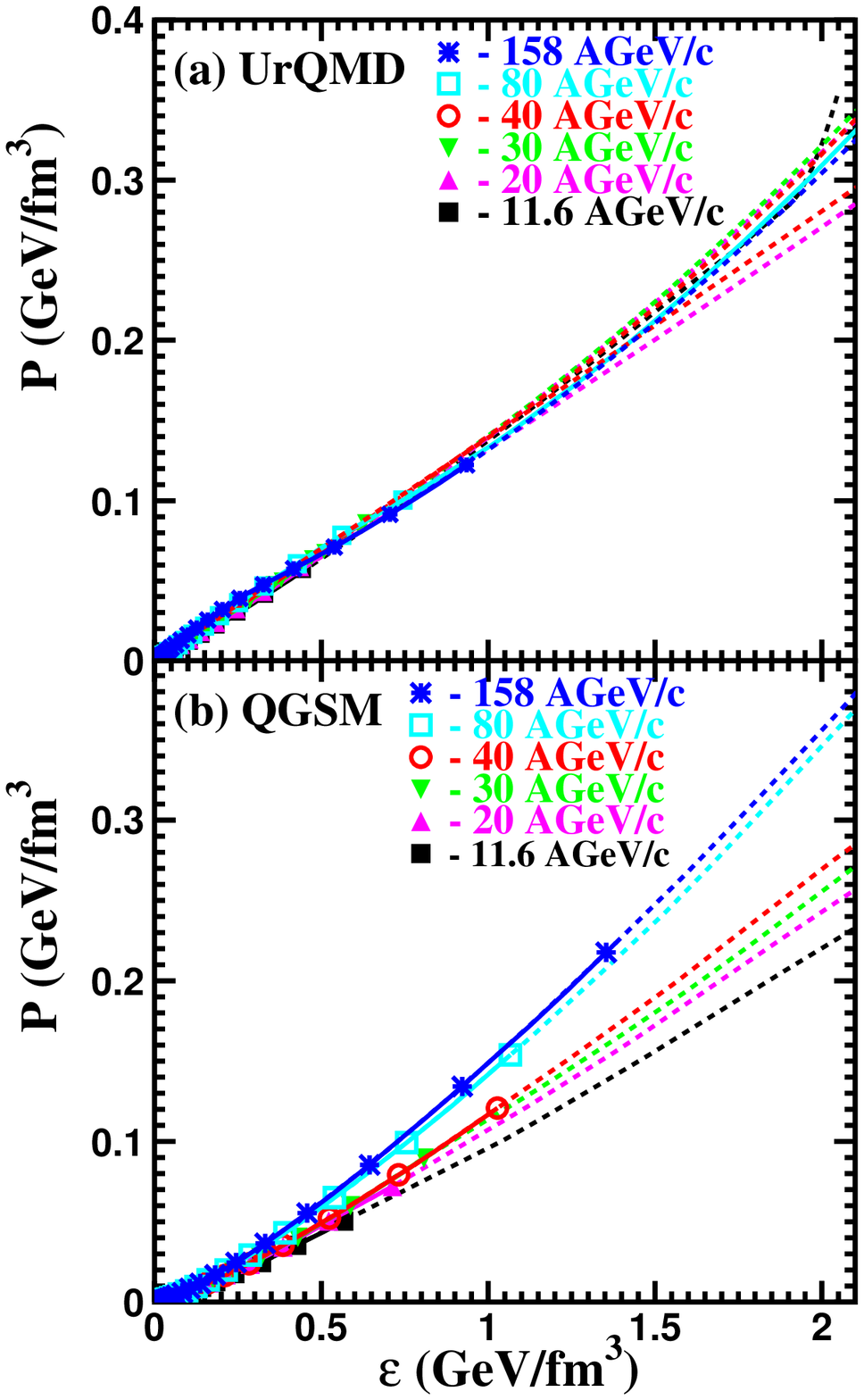,width=75mm}
\caption{
Equation of state: microscopic pressure $P$ vs. the energy density 
$\varepsilon$.
}
\label{fig4}
\end{minipage}
\end{figure}

Isentropic expansion of relativistic fluid is one of the main
postulates of Landau hydrodynamic theory \cite{La53} of multiparticle
production. As can be seen in Fig.~\ref{fig3} the entropy per 
baryon ratio is nearly conserved in the equilibrium phase of the 
expansion within the 5\% accuracy limit.
The entropy densities $s$ obtained for the cell
in both models are very close to each other, but, because of the
difference in net-baryon sector, the ratio $s/\rho_{\rm B}$ in
UrQMD is about 15--20\% larger than that in QGSM.

Any hydrodynamic model relies on the equation of state, which links
the pressure of the system to its energy density. Otherwise, the 
system of hydrodynamic equations is incomplete. The corresponding 
plot with microscopic pressures $P_{\rm mic}(\varepsilon)$ is 
presented in Fig.~\ref{fig4}. For both models the shapes of the
distributions are very close to linear for all energies in question. 
Thus the EOS has a rather simple form
\beq \ds
P(\varepsilon) = c_s^2 \varepsilon\ ,
\label{eq1}
\eeq
where the sonic velocity in the medium $c_s = (dP/d\varepsilon)^{1/2}$
is fully determined by the slopes of the distributions 
$P(\varepsilon)$. To account for possible deviations from a straight
line behavior the slopes of the functions $P$ versus $\varepsilon$ 
were averaged over the whole period of the equilibrated phase. For the 
UrQMD calculations the velocity of sound increases from 0.13 at 
$E_{\rm lab} = 11.6$\,AGeV to 0.146 at $E_{\rm lab} = 158$\,AGeV. It 
saturates at $c_s^2 \approx 0.15$ for RHIC energies \cite{urqmd_equil}.
That corresponds to change of the nuclear compressibility from 
140\,MeV\,(AGS) to 200\,MeV\,(SPS and RHIC). 
In QGSM calculations the averaged sound velocity is about 0.015 units 
smaller. Note that due to the averaging over time, respectively energy 
density, these values are lower the maximal values for $c_s^2$ 
that are reached in the corresponding reactions. Both models
indicate that at the energy around $E_{\rm lab} = 40$\,AGeV the 
slope of the $c_s^2 (\sqrt{s})$ distribution is changing, and the 
velocity of sound becomes less sensitive to rising bombarding energy.

\begin{figure}[htb]
\begin{minipage}[t]{70mm}
\epsfig{file=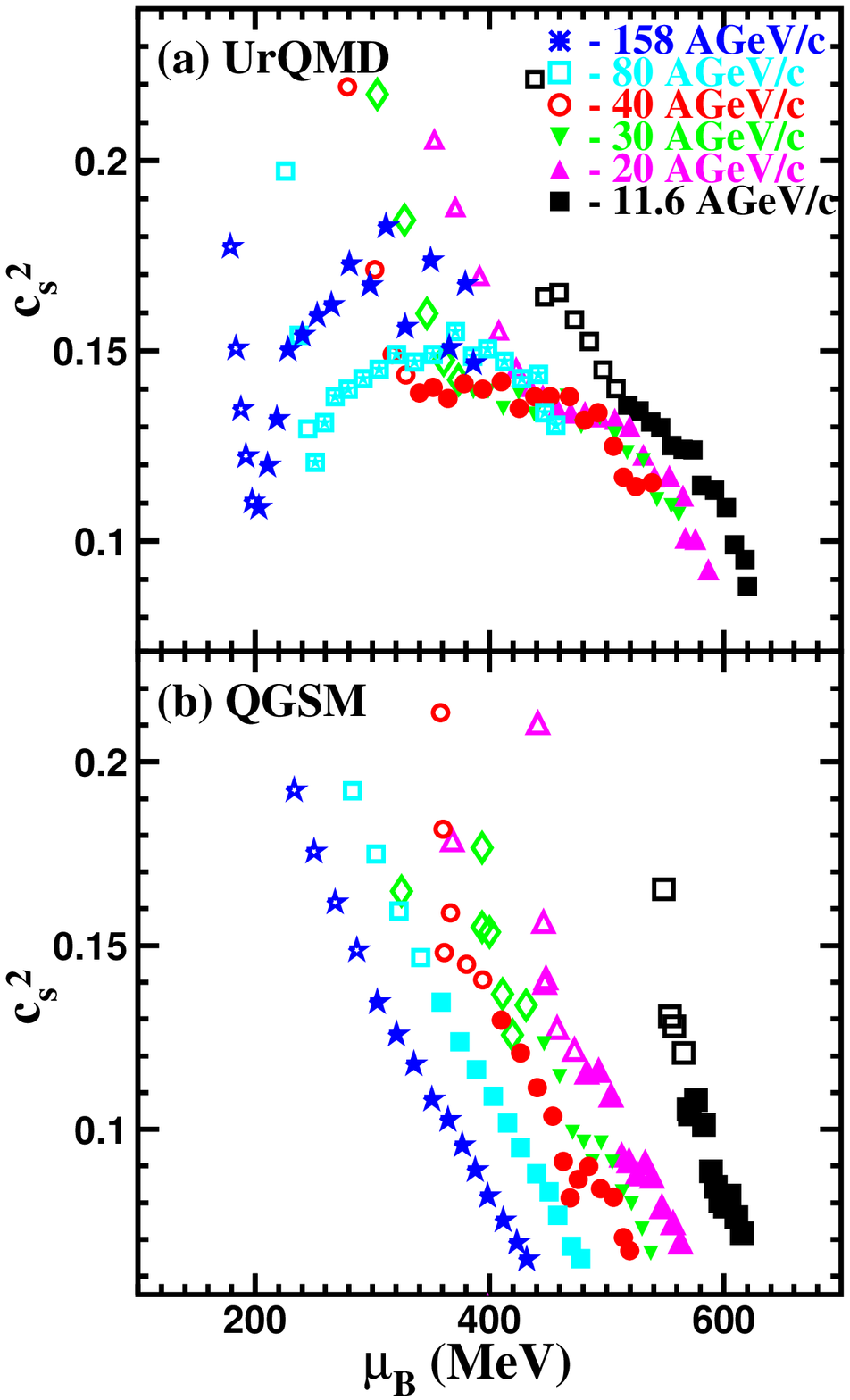,width=75mm}
\caption{
The sound velocity $c_s^2$ in the central cell of volume
$V=125$\,fm$^3$ as a function of baryon chemical potential 
$\mu_{\rm B}$. 
}
\label{fig5}
\end{minipage}
\hspace{\fill}
\begin{minipage}[t]{70mm}
\epsfig{file=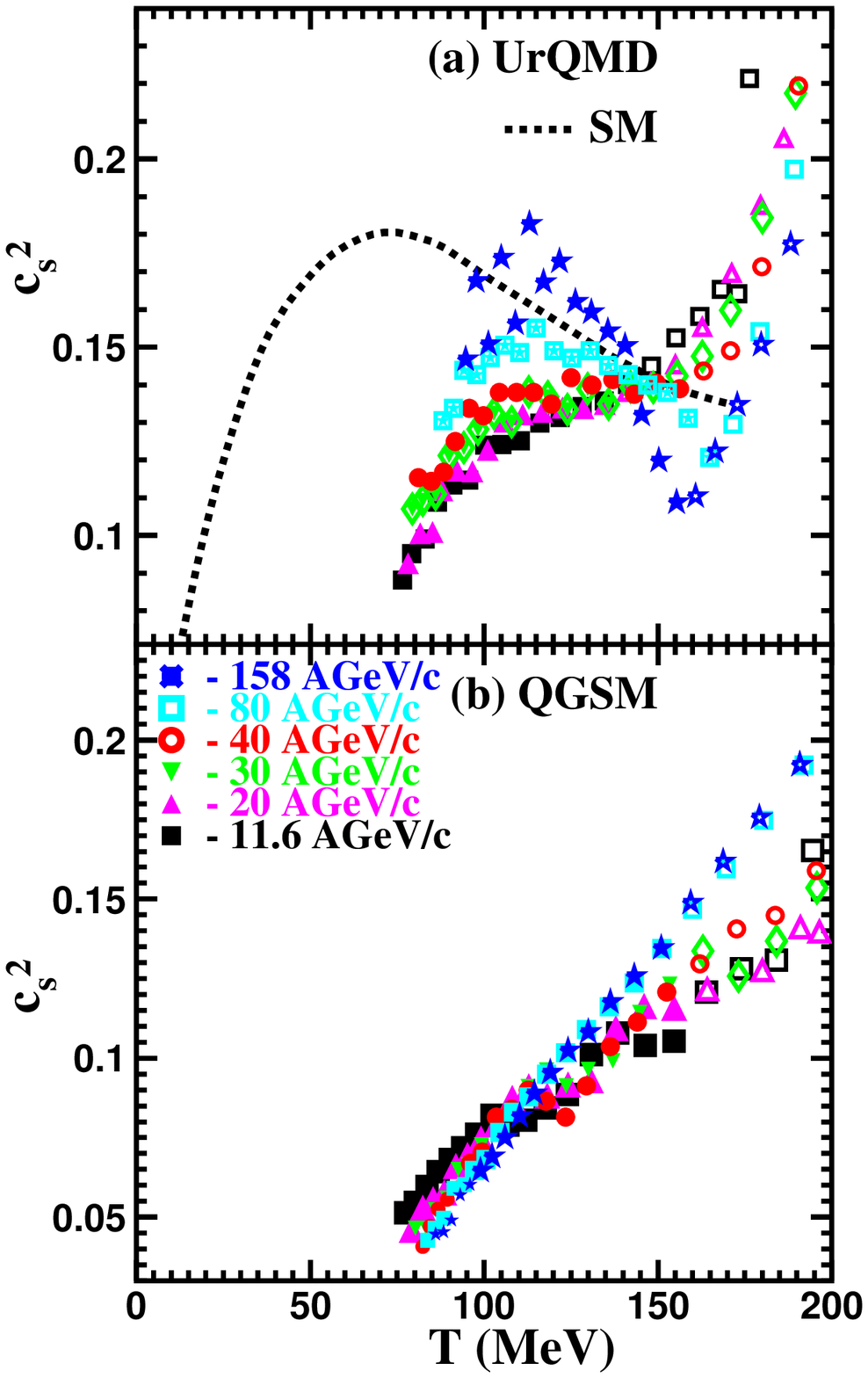,width=75mm}
\caption{
Temperature dependence of the sound velocity. Dashed line 
corresponds to calculations within Hagedorn model of ideal
hadron gas.
}
\label{fig6}
\end{minipage}
\end{figure}

Figure \ref{fig5} shows the dependence of the $c_s^2$ on the baryon
chemical potential $\mu_{\rm B}$. For three bombarding energies,
$E_{\rm lab} = 20$\,AGeV, 30\,AGeV, and 40\,AGeV, the
functions $c_s^2(\mu_{\rm B})$ are close to each other. In QGSM
calculations $c_s^2$ depends linearly on $\mu_{\rm B}$ and the
slope $c_s^2 / \mu_{\rm B}$ is unique for all reactions. In UrQMD the
picture is more complex. For the late stages of system evolution the
slopes of all distributions are also similar, but for energies of
$E_{\rm lab} \geq 40$\,AGeV one sees the rise of the sound 
velocity at the beginning of the equilibration, plateau, and the 
falloff. This can be taken as indication of the role of heavy 
resonances, because their fraction is presented in the particle 
spectrum at the early period and disappeared completely at the end. 
These resonances are rare at $E_{\rm lab} \leq 20$\,AGeV, and 
distributions $c_s^2(\mu_{\rm B})$ obtained in both models are quite 
similar.

The obtained EOS is soft, because for the ultrarelativistic 
gas of light particles the sonic speed is $c_s = 1/\sqrt{3}$. But the 
presence of resonances in particle spectrum generates the decrease 
\cite{Shur72} of the $c_s$. Employing the empirical dependence 
$ \ds \rho (m) \propto m^{\alpha^\prime} \ ,\
2 \leq \alpha^\prime \leq 3$ \cite{Hag65},
where  $\rho(m)\, dm$ denotes the number of resonances with masses from
$m$ to $m + dm$, one arrives to the equation of state in the form
\cite{Shur72}
\beq
\ds
\varepsilon = (\alpha^\prime + 4)\, P \ ,
\label{eq2}
\eeq
i.e., $\frac{1}{7} \leq c_s^2 \leq \frac{1}{6}$. This result is 
reproduced in microscopic models. 
Note that PHENIX collaboration reported the value 
$c_s \approx 0.35 \pm 0.05$ \cite{PHENIX_Cs}, i.e., $c_s^2 \approx 
0.12 \pm 0.3$, for Au+Au collisions at top RHIC energy $\sqrt{s} 
= 200$\,AGeV. This value is close to our results and also 
implies rather soft effective EOS.  

Temperature dependence of the sonic speed $c_s^2(T)$ is depicted 
in Fig.~\ref{fig6} together with the EOS calculated in \cite{CFC05} 
within the Hagedorn model with $\mu = 0$. For $E_{\rm lab} = 80$\,AGeV 
and 160\,AGeV the UrQMD data exhibit a falloff in $c_s^2(T)$ at 
$T \geq 120$\,MeV in accord with the Hagedorn model. This decrease
is assigned to heavy resonances, because neither the UrQMD 
calculations at lower energies nor the QGSM calculations without the 
heavy resonances reveal the negative slope in the equation of state 
$c_s^2(T)$. Below $T = 100$\,MeV both microscopic models indicate 
rapid drop of the sound velocity that occurs much earlier compared to
that of the Hagedorn model.

\begin{figure}[htb]
\begin{minipage}[t]{70mm}
\epsfig{file=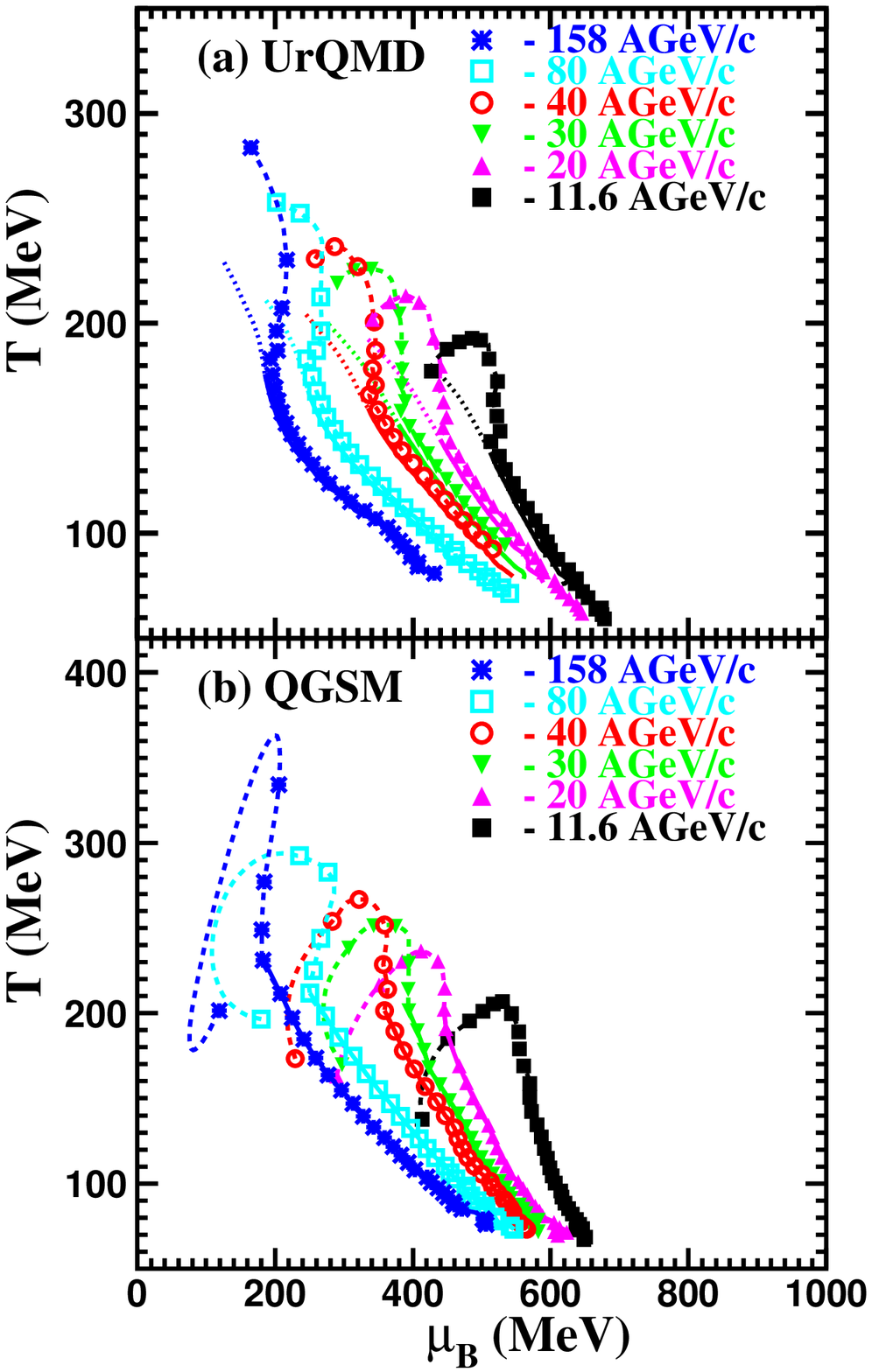,width=75mm}
\caption{
Temperature $T$ vs. baryon chemical potential $\mu_{\rm B}$.
}
\label{fig7}
\end{minipage}
\hspace{\fill}
\begin{minipage}[t]{70mm}
\epsfig{file=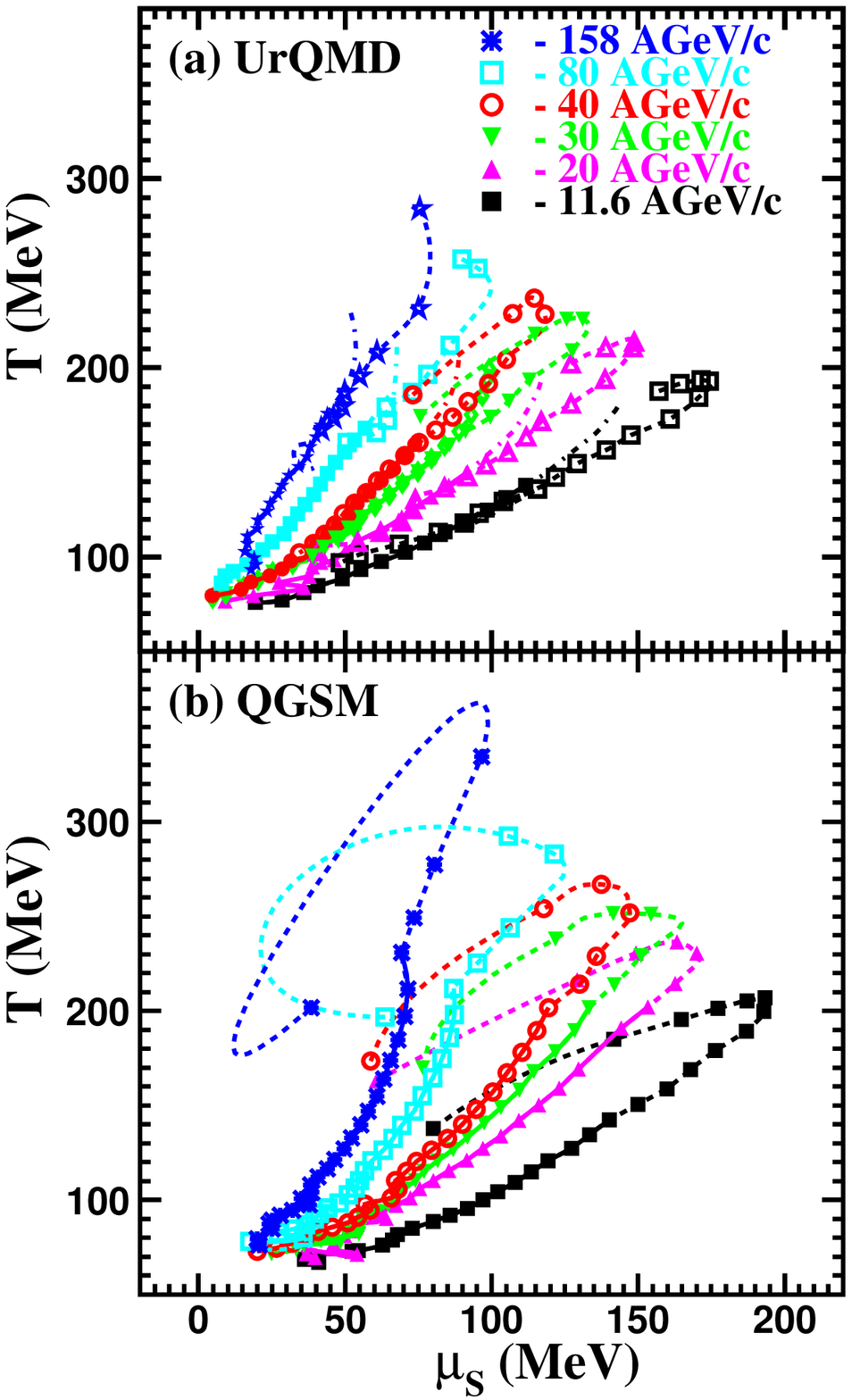,width=75mm}
\caption{
Temperature $T$ vs. strangeness chemical potential $\mu_{\rm S}$.
}
\label{fig8}
\end{minipage}
\end{figure}

In the modified analysis the central cell was further subdivided into 
the smaller ones embedded one into another.  If the $\varepsilon$ of 
the inner cell is not the same (within the 5\% limit of accuracy) as 
the energy density of the outer one, the SM analysis of the 
thermodynamic conditions is performed for the inner cell, otherwise 
the outer cell becomes a new test volume. This permits one to follow 
the expansion of the area with uniformly distributed energy.
EOS in the $T-\mu_{\rm B}$ plane is shown in Fig.~\ref{fig7}.
Symbols and dashed lines show the evolution of these quantities in a 
cell of instantly increasing volume ($V_{\rm init} = 0.125$\,fm$^3$), 
whereas dotted (upper plot) and full (both plots) lines are 
related to calculations with the fixed volume $V = 125$\,fm$^3$. 
The transition to equilibrium proceeds quite smoothly if the
analysis is performed for the fixed cell. In contrast, in the area 
with uniformly distributed energy the transition is characterized by 
a kink distinctly seen in each of the phase diagrams in both 
microscopic models. The effect, which takes place along the lines of 
the constant entropy per baryon, is caused by the significant 
reduction of the number of processes going via the formation and 
fragmentation of strings, i.e., chemical freeze-out.
The observed phenomenon can easily mimic the signature of the QCD phase 
transition in the $T$-$\mu_{\rm B}$ plane.
Evolution of strangeness chemical potential $\mu _{\rm S}$ with $T$
in the fixed volume and non-fixed volume is displayed in
Fig.~\ref{fig8}. As in Fig.~\ref{fig7}, all systems develop 
kinks in the $T(\mu_{\rm S})$ distributions precisely at the moment of
transition from nonequilibrium to equilibrium phase. Both baryon 
density and strangeness density are decreasing in the test volume,
however, the baryon chemical potential increases with time, whereas 
the strangeness one drops. The evolution of the $\mu _{\rm S}$ and 
$\mu _{\rm B}$ with $T$ proceeds quasilinearly, thus reducing the 
deviations, caused by nonzero chemical potentials, of the functions
$\varepsilon(T)$ and $s(T)$ from the ideal gas behavior at $\mu = 0$.

\section{Conclusions}
\label{sec3}

In summary, both microscopic models favor the formation of the 
equilibrated matter for a period of about 10\,fm/$c$ for all reactions
in question. During this period the matter in the 
central cell expands with constant entropy per
baryon. The equation of state can be approximated by a simple 
linear dependence $P = a(\sqrt{s}) \varepsilon$, where the square of
the speed of sound $c_s^2 = a(\sqrt{s})$ varies from 0.13 (AGS) to 
0.15 (SPS) in the UrQMD calculations and from 0.11 (AGS) to 0.14
(SPS) in the QGSM ones. 

Heavy resonances are responsible 
for negative slope in $c_s^2 (T)$ at $T \geq 100$\,MeV in accord with
the predictions of Hagedorn model of hadron resonance gas. At lower
temperatures both microscopic models indicate a rapid drop of the 
sonic speed in stark contrast with the
Hagedorn model calculations with zero chemical potential.
 
Study of the expanding area of isotropically distributed energy
reveals that the relaxation to equilibrium in this dynamic region
proceeds at the same rate as in the case of the fixed-size cell.
However, here both microscopic models 
unambiguously show the presence of a kink in the $T$-$\mu_{\rm B}$
phase diagrams. The higher the collision energy, the earlier the
kink formation. Its origin is linked to the freeze-out of inelastic 
reactions in the considered area.

{\it Acknowledgments\/.} This work was supported by the Norwegian
Research Council (NFR) under contract no. 185664/V30, by the DFG and 
the BMBF.



%
%

\end{document}